\documentclass[aps,prd,nofootinbib,preprint,tightenlines,longbibliography]{revtex4-1}
\pdfoutput=1

\usepackage{todonotes}
\usepackage{epsfig}
\usepackage{braket}
\usepackage{graphicx}
\usepackage{amsmath}
\usepackage{amsthm}
\usepackage{amssymb}
\usepackage{amsfonts}
\usepackage{mathbbol}
\usepackage{bm}
\usepackage{xcolor}
\usepackage{mathrsfs}
\usepackage{caption}
\usepackage{float}
\usepackage[pdftex,breaklinks,colorlinks,
citecolor=blue,
urlcolor=blue]{hyperref}

\def\down{\mathord{\downarrow}}
\def\up{\mathord{\uparrow}}
\def\diag{\mathop{\rm diag}}

\newcommand{\be}{\begin{equation}}
\newcommand{\ee}{\end{equation}}
\newcommand{\bea}{\begin{eqnarray}}
\newcommand{\eea}{\end{eqnarray}}
\def\bml{\begin{subequations}}
\def\blea{\bml\begin{eqnarray}}
\def\eml{\end{subequations}}
\def\elea{\end{eqnarray}\eml}

\def\bx{\mathbf{x}}
\def\bk{\mathbf{k}}

\def\mpl{m_{\text{Planck}}}
\def\Re{\operatorname{Re}}
\def\Im{\operatorname{Im}}

\def\Ndiff{N^{\text{diff}}}
\def\rhodiff{\rho^{\text{diff}}}

\begin{document}
\title{Energy conditions allow eternal inflation}

\author{Eleni-Alexandra Kontou}
\affiliation{Department of Mathematics, University of York,
  Heslington, York YO10 5DD, United Kingdom}
\affiliation{Department of Physics, College of the Holy Cross, Worcester, Massachusetts 01610, USA}
\author{Ken D. Olum}
\affiliation{Institute of Cosmology, Department of Physics and Astronomy, Tufts University, Medford, MA 02155, USA}

\begin{abstract}

Eternal inflation requires upward fluctuations of the energy in a
Hubble volume, which appear to violate the energy conditions.  In
particular, a scalar field in an inflating spacetime should obey the
averaged null energy condition, which seems to rule out eternal
inflation.  Here we show how eternal inflation is possible when energy
conditions (even the null energy condition) are obeyed.  The critical
point is that energy conditions restrict the evolution of any single
quantum state, while the process of eternal inflation involves
repeatedly selecting a subsector of the previous state, so there is no
single state where the conditions are violated.

\end{abstract}

\maketitle

\section{Introduction} \label{sec:introduction}
\subsection{Eternal inflation and the null energy condition}

\begin{figure}
	\centering
	\includegraphics[scale=0.5]{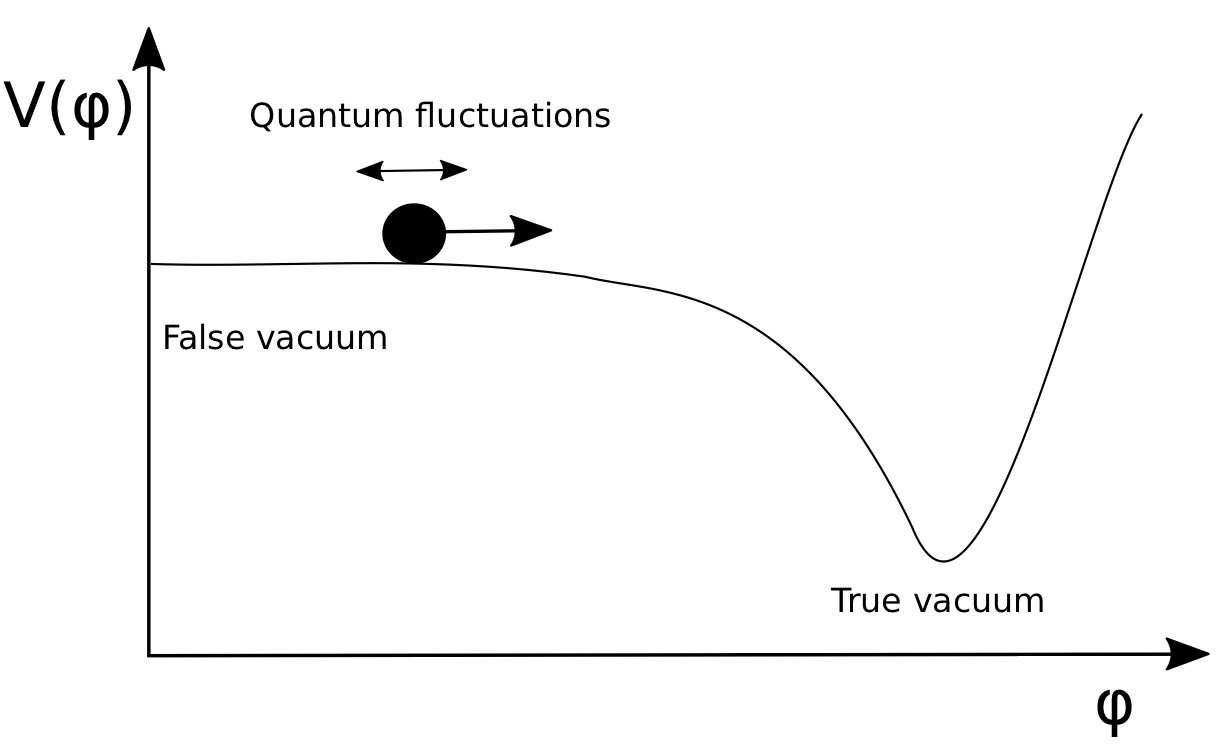}
	\caption{An example of a slow-roll inflaton potential. The field classically rolls down to the true vacuum region. However, quantum fluctuations can drive the field up the slope, increasing the rate of expansion.}
        \label{fig:potential}
\end{figure}

Inflation is a period of quasi-exponential expansion driven by the
potential energy of some inflaton scalar field $\phi$
\cite{Guth:1980zm}.  Classically, the value of $\phi$ rolls slowly
down a mildly sloped potential, giving time for many $e$-foldings of
expansion of the universe.  However, in most scenarios
\cite{Guth:2007ng}, quantum fluctuations sometimes drive $\phi$ up the
potential in a particular Hubble volume, causing an increase in the
expansion rate there.  See Fig.~\ref{fig:potential}.
While such fluctuations are rare, the regions in which they occur then
expand faster than others.  Each such volume becomes many volumes,
each of which may experience another upward fluctuation, causing
inflation to persist eternally.

Eternal inflation, however, seems to require violation of energy
conditions, which are restrictions on the stress-energy tensor $T_{\mu\nu}$ of
quantum fields.  (See \cite{Kontou:2020bta} for a recent review).  We will be particularly concerned with the null
energy condition (NEC), which requires that
\be
T_{\mu\nu} \ell^\mu \ell^\nu\ge0
\ee
for all null vectors $\ell$.  If $T_{\mu\nu}$ has the form of a
perfect fluid with energy density $\rho$ and isotropic pressure $P$ in
its rest frame, NEC takes the form
\be
\rho+P \ge0\,.
\ee

Borde and Vilenkin \cite{Borde:1997pp}, Winitzki
\cite{Winitzki:2001fc}, and Vachaspati \cite{Vachaspati:2003de} have
argued that eternal inflation requires violations of the Null Energy
Condition (NEC), as follows. In a Friedman-Robertson-Walker universe
with metric
\be
ds^2=dt^2-a(t)^2 d\vec{x}^2 \,,
\ee
the equation governing the Hubble expansion, $H=\dot{a}/a$, is
\be
\dot{H}=-4\pi G (\rho+P)+\frac{k}{a^2} \,.
\ee
In an inflating spacetime, we can ignore the curvature term so set
$k=0$. Eternal inflation requires an increase in the expansion rate,
meaning $\dot{H}>0$, so $\rho+P <0$,
and the NEC is violated. 

\subsection{The averaged null energy condition}

The need for NEC violation is not in itself a big problem, since
quantum fields can easily violate NEC or any pointwise energy
condition \cite{Epstein:1965zza}. The problem becomes much worse when we consider the
averaged null energy condition (ANEC),
\be
\label{eqn:ANEC}
\int_\gamma \, d\lambda \, T_{\mu \nu} \ell^\mu \ell^\nu \geq 0 \,,
\ee
where the integral is taken over a complete geodesic $\gamma$.  ANEC
is much harder to violate than NEC.  We proved \cite{Kontou:2015yha}
that ANEC is obeyed by a quantum scalar field on a curved background
generated by a classical field.  Here we can consider the background
inflation to be given by a classical inflaton field and fluctuations
in that field to be quantum-mechanical.\footnote{Our proof was for a
  massless scalar, but in inflation we have a scalar field with a
  potential.  However the potential is nearly flat and level, so the
  corrections to the previous analysis are small.  There are some
  other technical restrictions, but they are all obeyed in the present
  case.}

We should note here that null geodesics are not past-complete in an inflating spacetime
\cite{Borde:2001nh}, so ANEC is not directly applicable.  Nevertheless
one can start with a spacetime that is de Sitter space near
$t=-\infty$ in the Robertson-Walker parameterization, and argue that
ANEC applies to the part of the geodesic starting at $t=-\infty$.

Reference~\cite{Borde:1997pp} shows that the ANEC integral during
eternal inflation can be written
\bea
\label{eqn:hintegral}
&&\int_\gamma \, d\lambda \, T_{\mu \nu} \ell^\mu \ell^\nu=-2\int dt \frac{\dot{H}}{a} \,.
\eea
This is an integral over $\dot H$ with weight factor $1/a$, which
falls exponentially with time.  Thus an increase in expansion rate
could obey ANEC if it was preceded by a decrease.  The decrease could
even be much smaller if it took place a while earlier.  But there does
not seem to be anything about the fluctuations of $\phi$ that would
require a decrease at all.  The model of fluctuating $\phi$ includes
scenarios in which fluctuations drive the $\phi$ monotonically up the hill in the potential.  Such
scenarios would violate ANEC, and there seems to be no reason why they
could not happen.  We will see that the solution to this problem lies
elsewhere.

There is a simple geometrical argument that shows the relationship
between ANEC and eternal inflation.  ANEC says, essentially, that
gravity does not allow defocusing of null geodesics, so there is no
``anti-lensing''.  Now imagine a converging spherical shell of
incoming geodesics infinitesimally inside the Hubble distance.  If a
fluctuation increases the expansion rate, the Hubble distance will be
smaller, so the sphere is outside the Hubble distance, and the
expansion of the universe carries the geodesics away.  This means that
the geodesics are now diverging and that requires ANEC violation.

\subsection{Quantum-mechanical expectation values}

One might argue that energy conditions restrict only the quantum
mechanical expectation value of $T_{\mu\nu}$ and this is
unproblematic.  Perhaps $\langle T_{\mu \nu}\rangle$ in inflation
indeed obeys NEC and ANEC, and the violations necessary for eternal
inflation take place only in some sector of this state, with other
sectors outweighing the problematic sector in $\langle T_{\mu
  \nu}\rangle$.  But this is not correct, because when a field obeys
an energy condition, it obeys it in every quantum state. This was
shown in \cite{Fewster:1999kr} where quantum inequalities are
reformulated from restrictions on expectation values to statements
about the positivity of certain operators.  Thus if the inflating
state can be written as a sum of sectors, $|\psi\rangle = \sum_i
\alpha_i|\psi_i\rangle$, every sector $|\psi_i\rangle$ must obey ANEC.
There cannot be any sector where ANEC is violated, allowing $H$ to
increase.

\subsection{A solution to the apparent paradox:  selection of sectors}

The right solution to this problem is slightly different.  It's not
that the eternally inflating state is a substate of the quantum state
during inflation.  Rather, eternal inflation requires (repeated)
selection of a substate that has a faster expansion rate.  Thus
rather than being a single quantum state to which energy conditions
and their consequences would apply, eternal inflation has one state at
early times and a different state later.  The later state is selected
from the much broader earlier state to be one that happens to have
faster expansion in a certain volume.  Since the faster expansion
arose from selection rather than just evolution of the earlier
stage, we cannot prove something about the final state using energy
conditions.

\subsection{A simple model for eternal inflation}
\label{sub:model}

In this paper, we will primarily use a slightly simplified model of
inflation that nevertheless supports eternality and exhibits the
problem that we are trying to solve.  Application to realistic
scenarios is discussed in Sec.~\ref{sec:realistic}.  We will consider
a massive scalar field $\phi$ evolving as if in a background de Sitter
space with (fixed) Hubble constant $H$.  We will allow $\phi$ to
change the metric, but we will not allow the changed metric to affect
the evolution of $\phi$.  We thus ignore the rolling of the inflaton
that leads to inflation being only quasi-de Sitter, and we also ignore
the gravitationally-mediated effect of fluctuations of $\phi$ on its
equation of motion.

This model is sufficient to lead to eternal inflation, because a
superhorizon fluctuation of $\phi$ leads to a slowly rolling energy
density $\rho = (1/2)m^2\phi^2$ that increases the expansion rate.
This is possible in our model because we do allow the energy of the
changed field to affect the metric, even though we do not allow the
resulting metric changes to affect $\phi$.

We can thus take $\phi$ to be given by quantum field theory in a fixed
de Sitter background.  When we discuss the expansion rate, we will
take the shape of spacetime to passively follow the instantaneous
value of the field $\phi$.  Thus we will not consider propagating
modes of gravity.  The gravitational field has no quantum mechanics of
its own but merely follows the quantum mechanics of $\phi$.  One might
call this approach ``passive quantum cosmology'' in analogy to the
passive quantum gravity of Ford and Wu \cite{Ford:2000vm}.  

This approach is very different from semiclassical gravity, where one
takes the quantum expectation value $\langle T_{\mu\nu} \rangle$ as
the source for gravitation.  In semiclassical gravity, it is not
possible to consider quantum mechanical alternatives for the shape of
spacetime, as we will do here.  Rather, the shape of spacetime is
uniquely determined by a single quantum mechanical average.

\subsection{Plan of paper}

In the rest of this paper we will carefully explain the idea above.
We start in Sec.~\ref{sec:analogies} with some simple models showing
how selection can find values of a quantity that would otherwise be
prohibited.  In Sec.~\ref{sec:class} we derive the classical field
equations for our model, and in Sec.~\ref{sec:squeezed} we quantize
the field assuming that it is in a general Gaussian state and examine
two particular states, the Bunch-Davies vacuum and a late time rapid
expansion state which is the one corresponding to the eternally
inflating part of spacetime.  In Sec.~\ref{sec:energy} we evolve this
(selected) late state backward in time and show that this specific
sector obeys NEC and thus ANEC.  In fact this rapidly-expanding sector
had even more rapid expansion earlier.  In Sec.~\ref{sec:realistic} we
show how these ideas extend to more usual inflationary models, and we
conclude in Sec.~\ref{sec:conclusion}.

\section{Some simple analogies}\label{sec:analogies}

\subsection{Momentum in a linear potential}

We begin with a simple analogy with ``uphill'' fluctuations.  Consider
a particle in a linear potential $V(x) = - f x$, where $f$ is a
constant.  Classically, the particle's momentum increases steadily,
$dp/dt = f$.  In quantum mechanics, Ehrenfest's theorem tells us that
$d\langle p\rangle/dt = f$.  By analogy with energy conditions, one
might call this a ``momentum condition'' saying that $\langle
p\rangle$ increases in every state.

We can write the momentum-space wave function $\psi(p)$.  The
Schr\"odinger equation is $\partial\psi/\partial t = ip^2 /(2m)\psi -f
\partial \psi/\partial p$ and the general solution is $\psi(p;t) =
\psi_0(p-ft)\exp(ip^2 t/(2m))$, where $\psi_0$ is the wave function at
time $t=0$.  Now let $\psi_0$ be a Gaussian centered around 0.
Suppose we measure the momentum at some time $t_1$.  The most likely
result will be $ft_1$, but there is a possibility that we will find
some $p_1<0$.  In such a case the momentum is now less than the
(average) momentum at $t = 0$, even though $\langle p\rangle$
increases in every state.  No measurement can be perfect, so our
measurement apparatus will leave the particle in some state, say a
Gaussian, $\psi(p;t_1)$ centered around $p_1$.  Looking backward, the
state we discovered had $\langle p\rangle= p_1-ft_1$ at $t = 0$.

If we measure the state again at some time $t_2>t_1$, there's a small
chance we will find $p_2<p_1$, and so on.  Repeated measurements can
lead to a decreasing $p$ even though classically $p$ always increases
and quantum mechanically $\langle p\rangle$ always
increases.

We have assumed here an ideal measurement that
  measures the momentum (approximately), without the measurement
  process itself injecting momentum into the system.  It is not clear
  whether such a measurement can really be made.  However, if we
  consider a free particle ($f=0$), then we can measure its momentum
  without affecting the system (other than the projection required for
  measurement), because momentum eigenstates are also eigenstates of the
  Hamiltonian.  In that case $\langle p\rangle$ is constant in every
  state but can decrease by ideal measurement with no momentum flowing
  from the measurement apparatus.

\subsection{Scalar field in flat space}
\label{sub:scalar}

Let us now consider a field theory example: a free massless scalar in flat space.
We will work in a periodic cubical box of side $L$ and volume $V =
L^3$.  We decompose the field in Fourier modes:
\be
\phi(\bx, t) = \frac1{L^{3/2}}\sum_{\bk} e^{i\bk\cdot\bx} \phi_k(t) \,.
\ee
Since $\phi$ is real, we must have $\phi_{-k}=\phi_k^*$.  We can
decompose it,
\be
\phi_k=\chi_k+i \xi_k \,,
\ee
where $\chi_k$ and $\xi_k$ are real fields.  We will concentrate here
on the $\chi_k$; the $\xi_k$ are entirely analogous.

Each mode is just a harmonic oscillator with Hamiltonian
\be
H_k = \frac12\left[p_k^2+k^2\chi_k^2\right]
\ee
where the momentum associated with $\chi_k$ is $p_k = \dot\chi_k$.
The classical equation of motion is
\be
\chi_k''+k^2\chi_k = 0\,.
\ee
We quantize the field as usual.  The ground state wave function is
just a product of wave functions for the individual modes, each of
which is a Gaussian
\be
\psi(\chi_k) = \left(\frac{k}{\pi}\right)^{1/4}e^{-k\chi_k^2/2}\,,
\ee
which has ground state energy $k/2$.  (Usually we would renormalize
by subtracting the ground state energies, but it does not matter here.)

Now consider some specific mode $\bk$ and suppose we measure $\chi_k$.
We might find $\chi_{k1} \gg (k)^{-1/2}$ and thus energy much larger
than $k/2$.  After measurement, we suppose that $\chi_k$ will be in a
Gaussian around $\chi_{k1}$, which evolves as a squeezed coherent
state \cite{Kumar:2000zi} whose center and width both oscillate with time.
Subsequent measurement of $\chi_k$ may find an even larger energy, so
again we conclude that repeated measurement sometimes raises the
energy of the system.

However, the system started in the ground state, which is an
eigenstate of energy, so the energy cannot fluctuate.  The increasing
energy in this case must be added by the measurement apparatus.
Without such a process, the energy would be fixed.  Thus this system
is a poor analogy to inflationary cosmology.  In that case,
decoherence may select out specific sectors of a wave function, but
there is no measurement apparatus that could add energy to the system.

\subsection{Small changes to $H$ allow large changes to $E$}
\label{sub:smallchanges}

The reason that, as we will see, decoherence can select modes with
different energies, is that the Hamiltonian for each mode changes over
time due to the expansion of the universe.  We will discuss this in
detail below, but first we'll give an example showing how a small change
to the Hamiltonian can lead to large (but rare) changes to an otherwise
conserved energy.

Consider a spin in a magnetic field pointing along the $z$ axis, so
that the Hamiltonian is $H_1 =
E(|\up\rangle\langle\up|-|\down\rangle\langle\down|)$.  At time $t =
0$, let us rotate the magnetic field by a tiny angle $\epsilon$, so
the Hamiltonian is\footnote{It's important that the change is rapid
  compared to the timescale set by the eigenvalue difference $2E$.
  Otherwise the adiabatic theorem would tell us that the system would
  remain in an eigenstate of energy.}
$H_2=E(|\up'\rangle\langle\up'|-|\down'\rangle\langle\down'|)$.
Prepare the initial state $|\down\rangle$, with energy $-E$.  At
$t=0$, this state will become
$\cos(\epsilon/2)|\down'\rangle+\sin(\epsilon/2)|\up'\rangle$.  We can
now measure the spin along the new magnetic field direction without
adding energy to the system.  With probability $\sin^2\epsilon$ we
will find energy $E$ even though our system was prepared with energy
$-E$.

It's not surprising that energy is not conserved when the Hamiltonian
changes.  But it's worth noting that the change in energy can be large
even when the change in the Hamiltonian is tiny.  The eigenvalues of $H_2-H_1$ are
$\pm2E\sin(\epsilon/2)\approx \epsilon E$ if $\epsilon\ll1$.  A tiny
change in the Hamiltonian can allow a large but rare change in the
energy of the system.

The analogy here is that the upward fluctuations needed for eternal
inflation are very rare, but the state when they happen is quite
different from the original inflating state.  The Hamiltonian for the
modes of the inflaton is time-dependent because of the expansion of
the universe.  For super-horizon modes, the time dependence is rapid
compared to the oscillation frequency of the mode.  Thus even though
each mode starts in its ground state, it is not in the ground state
later.  Decoherence will then select a state in which each mode has
its value distributed in a narrow range.  If those ranges happen to be
displaced up the potential, the result will be an increase in the
expansion rate, even though ANEC is never violated in any quantum
state.

\section{Classical field equations}
\label{sec:class}

We now return to the simple eternally inflating model of
Sec.~\ref{sub:model} to understand how upward fluctuations can be
compatible with energy conditions.  Inflaton fluctuations on a
classical FRW background have been discussed by various authors, see
\cite{Mukhanov:1990me} for one of the first descriptions of both
classical and quantum perturbation and \cite{Riotto:2002yw} for a
pedagogical introduction. We begin by analyzing the classical
evolution of the massive scalar field $\phi$ in the de Sitter
background.

As in Sec.~\ref{sub:scalar}, we work in a box of comoving side length $L$ and volume $V=L^3$.  The field $\phi(\bx, t)$ can then be decomposed in Fourier
components,
\be
\label{eqn:modes}
\phi(\bx, t) = \frac{1}{L^{3/2}} \sum_{\bk} e^{i\bk\cdot\bx} \phi_k(t) \,.
\ee
with $\phi_{-k}=\phi_k^*$.

The Lagrangian density for a massive scalar is
\be
\mathcal{L} = \frac12\left[\dot\phi^2-(\nabla\phi)^2-m^2\phi^2\right] \,,
\ee
and we can write the total Lagrangian $\sum L_k$ where
\be
L_k = \frac {a^3}2\left[|\dot\phi_k|^2-\left(\frac{k^2}{a^2}+m^2\right)|\phi_k^2|\right] \,,
\ee
where $a(t)$ is the scale factor.
The canonical momentum is therefore $p_k = a^3 \dot\phi_k^*$ and the
Hamiltonian for the mode is
\be
\label{eqn:hamdot}
H_k = \frac12\left[\frac{|p_k|^2}{a^3}+a^3\left(\frac{k^2}{a^2}+m^2\right)|\phi_k|^2\right] \,.
\ee
The equation of motion is
\be\label{eqn:EOM1}
\ddot\phi_k+3H\dot\phi_k+\left[\frac{k^2}{a^2}+m^2\right]\phi_k = 0\,,
\ee
representing a damped oscillator with varying frequency.

In this simple system where changes to the metric due to $\phi$ are
prohibited from affecting the equation of motion, and with a quadratic
potential only, the modes evolve independently, which allows us to
study evolution of each $\bk$ in isolation.  In a realistic system,
there would be a more complicated potential and also gravitational
coupling between the modes.  In fact we will rely later on the
existence of some such couplings to provide decoherence, but we will
assume that they are small enough that we can treat the evolution of
the modes separately.

To solve Eq.~(\ref{eqn:EOM1}), we first go to conformal time $\tau$,
defined by $a d\tau=dt$, and write conformal time derivatives $\phi' =
d\phi/d\tau = a \dot\phi$. The action $S =\int L dt$ should be
unchanged, so the Lagrangian should be multiplied by $dt/d\tau =a$,
\be
L_k = \frac{a^2}2
\left[|\phi_k'|^2-\left(k^2+a^2m^2\right)|\phi_k|^2\right] \,.
\ee
Then we define a new field $\sigma_k = a \phi_k$, in terms of which
\be
L_k = \frac{1}2
\left[|\sigma_k'|^2-\frac{a'}{a}(\sigma_k^*\sigma_k'+\sigma_k {\sigma_k^*}')
-\left(k^2+a^2m^2-\frac{a'^2}{a^2}\right)|\sigma_k|^2\right] \,.
\ee
We're free to add the total derivative $(d/d\tau)(\sigma_k^2 a'/a)$ giving
\be
L_k = \frac{1}2
\left[\sigma_k'^2 -\left(k^2+a^2m^2-\frac{a''}{a}\right)\sigma_k^2\right]
\ee
representing a varying-frequency but undamped oscillator.
The equation of motion is
\be\label{eqn:EOMsigma}
\sigma_k'' + K(\tau) \sigma_k = 0 \,,
\ee
with
\be\label{eqn:EOMK}
K(\tau) = k^2+a^2m^2-\frac{a''}{a} \,.
\ee

We will work in de Sitter space, $a(t) = a_0e^{Ht}$, where $H$ is the
unchanging physical Hubble constant, so $\tau =-(a_0/H)e^{-Ht}$, and
runs from $-\infty$ up to 0.  We find $a(\tau)= -1/(H\tau)$, $a''/a =
2/\tau^2$ and
\be
\label{eqn:Kexpr}
K(\tau) = k^2 + \frac{1}{\tau^2}\left[\frac{m^2}{H^2}-2\right] \,.
\ee
The general solution to Eqs.~(\ref{eqn:EOMsigma},\ref{eqn:Kexpr}) is
\be\label{eqn:classsolJY}
\sigma_k = \sqrt{-\tau}\left[c_J J_\nu(-k\tau) + c_Y
Y_\nu(-k\tau)\right]
\ee
with
\be
\label{eqn:nu}
\nu^2 = \frac94 -\frac{m^2}{H^2}\,.
\ee
We are interested in $m \ll H$, so we can write
\be
\nu =\frac32 - \epsilon \,,
\ee
with
\be\label{eqn:epsilon}
\epsilon = \frac32-\sqrt{\frac94 -\frac{m^2}{H^2}}
\approx \frac{m^2}{3H^2} \,.
\ee
The coefficients in Eqs.~\eqref{eqn:classsolJY} are complex, but since
$\sigma_k=\sigma_{-k}^*$, we require $c_J(-k) =c_J(k)^*$ and likewise
for $c_Y$.  Thus there are 4 real degrees of freedom for the system
including wave vectors $k$ and $-k$.

Including exponential factors and both $\sigma_k$ and $\sigma_{-k}$,
the contribution to $\phi$ from Eq.~(\ref{eqn:classsolJY}) is

\be\label{eqn:classsolJY-}
e^{i\bk\cdot \bx} \phi_{k}(t)+e^{-i\bk\cdot \bx}
  \phi_{-k}(t)=
\frac{2\sqrt{-\tau}}{a} \left[|c_J|\cos(kx+\delta_J)J_\nu(-k\tau)
+ |c_Y|\cos(kx+\delta_Y) Y_\nu(-k\tau)\right]
\ee
with $\delta_J = \arg c_J$ and similarly for $Y$.

At early times ($a\to0;t\to-\infty$), the solutions to
Eq.~(\ref{eqn:EOMsigma}) are simply oscillations.  Using the
large-argument approximation to $J_\nu$ and $Y_\nu$ in
Eq.~(\ref{eqn:classsolJY-}) gives
\be\label{eqn:classsolearly}
\frac{1}{a}
\sqrt{\frac{8}{\pi k}}\bigg[|c_J|\cos(kx+\delta_J)\cos(k\tau +\delta_\nu)
- |c_Y|\cos(kx+\delta_Y)\sin(k\tau + \delta_\nu)\bigg]
\ee
where $\delta_\nu =(\pi/2)(\nu+1/2)$.  This is a superposition of
two damped standing waves, and we can write it also as a superposition of
damped traveling waves going in opposite directions.

At late times ($a\to\infty; \tau\to0^-$), we use the small-argument
approximation in Eq.~(\ref{eqn:classsolJY-}) to get
\be
\frac{(-k\tau)^{\nu+1/2}}{2^\nu\Gamma(\nu+1) a \sqrt{k}}
|c_J|\cos(kx+\delta_J)
+ \frac{2^\nu\Gamma(\nu)(-k\tau)^{1/2-\nu}}{a \sqrt{k}}| c_Y|\cos(kx+\delta_Y) \,.
\ee
In the first term, 
\be
\sigma\sim (-\tau)^{\nu+1/2}
\ee
so
\be
\phi\sim (-\tau)^{\nu+3/2} \sim (-\tau)^{3+\epsilon}
\ee
which declines rapidly.  But in the second term
\be
\sigma\sim (-\tau)^{1/2-\nu}
\ee
so
\be\label{eqn:slowmode}
\phi\sim (-\tau)^{3/2-\nu} =
(-\tau)^\epsilon \sim a^{-\epsilon}\sim e^{-\epsilon Ht}\sim e^{-m^2t/(3 H)}\,,
\ee
which is a slow-roll mode.  In the massless case it would be frozen.
This is just what you would get by starting from Eq.~\eqref{eqn:EOM1},
ignoring the second-derivative term to find the slow-roll solution,
and going to late times where $a\gg k/m$.

The conventional slow-roll parameters are
\be
\varepsilon = \frac1{16\pi G}\left(\frac{U'}{U}\right)^2 \,,
\qquad
\eta = \frac1{8\pi G}\left(\frac{U''}{U}\right)\,.
\ee
where $U$ is the scalar field potential.  In our case, $U' = m^2\phi$,
but for $U$ we should use instead the effective potential of the
cosmological constant that is driving the underlying expansion,
$U=3H^2/(8\pi G)$, giving
\be
\label{eqn:slowroll}
\varepsilon = \frac{4\pi\phi^2}{\mpl^2}\epsilon^2 \,, \qquad
\eta = \epsilon\qquad
\ee
for $\epsilon\ll1$.  Thus our solution qualifies as slow roll for
sub-Planckian field values.

Note that at late times there are no right-going and left-going
modes.  Propagating modes appear in Eq.~(\ref{eqn:classsolearly})
through particular choices of the $J$ and $Y$ parameters.  But the $J$
sector decreases rapidly once the mode crosses the horizon, leaving
only a slowly decreasing sinusoidal in the $Y$ sector.

If we discover the field with some value $\phi_1$ at some time
$\tau_1$, the simplest possibility would be that it is in the slow-roll mode
with a pure $Y_\nu$ solution,
\be\label{eqn:phi1classical}
\phi_1 = \frac{C_Y \sqrt{-\tau} Y_\nu(-k\tau)}{a} =
-\frac{C_YH\Gamma(\nu)}{\pi}\left(\frac{2}{k_1}\right)^{3/2}\left(-\frac{2}{k\tau_1}\right)^\epsilon \,.
\ee
But we cannot rule out some admixture
of the rapidly decaying mode, unless we measure $\dot\phi$ very
carefully.

Extending Eq.~(\ref{eqn:phi1classical}) to early times, we find the
solution
\be
\phi = \frac{C_Y}{a}\sqrt{\frac{2}{\pi k}}\cos(-k\tau +\delta)
= -\frac{\sqrt{\pi}k\phi_1}{2H\Gamma(\nu)a}\left(-\frac{k\tau_1}{2}\right)^\epsilon
\cos(-k\tau +\delta) \,.
\ee
Thus we know that such a state at early times has an oscillation at
least as large as that, and there could have been a much larger
oscillation (with a different phase) that has decayed faster.

\section{Quantum mechanics}
\label{sec:squeezed}

We quantize the field $\phi$ in the Schr\"odinger picture.  Formally
we will write a wave functional $\Psi[\phi]$ obeying $i \partial
\Psi/\partial t=H \Psi$.  But to have a tractable situation, we will
consider only $\Psi$ that are products of terms for the various $\bk$,
$\Psi[\phi]=\prod\Psi_k(\phi_k)$.  We will furthermore restrict our
consideration to cases where $\Psi_k$ is a product of wavefunctions
operating on the real and imaginary parts of $\phi_k$.  We decompose
\be
\phi_k=\chi_k+i \xi_k \,,
\ee
 as in Sec.~\ref{sub:scalar}, and require
 $\Psi_k(\phi_k)=\psi_k(\chi_k)\zeta(\xi_k)$.  We discuss below why
 this is sufficient for our purposes.

We let momentum operator corresponding to $\chi_k$ be $p_k = -i
\partial/\partial\chi_k$.  The time-dependent Schr\"odinger equation
for $\psi_k$ is then (see Eq.~(\ref{eqn:hamdot})),
\be
\label{eqn:schr}
i \frac{\partial \psi_k}{\partial t}=H_k \psi_k
=\frac12\left[-\frac1{a^3}\frac{\partial^2}{\partial\chi_k^2}
+a^3\left(\frac{k^2}{a^2}+m^2\right)\chi_k^2\right]\psi_k \,.
\ee

We will look for solutions where $\psi_k(\chi_k)$ has a Gaussian form.
Ehrenfest's theorem tells us that the expectation values of the
position and momentum satisfy the classical equations of motion.  So
let us choose some classical solution $\Phi_k(t)$ and
$P_k(t)=a^3\dot\Phi_k(t)$, and look for $\psi_k(\chi_k)$ in the form
\be
\label{eqn:state}
\psi_k(\chi_k,t)=N(t) \exp{\left[-\frac{A_k(t)}{2}(\chi_k-\Phi_k(t))^2+iP_k(t)\chi_k+i\theta_k(t)\right]} \,,
\ee
where $N_k(t)$, $A_k(t)$, and $\theta_k(t)$ are functions to be
determined.  Here $N_k(t)$ and $\theta_k(t)$ are real, but $A_k(t)$
can be complex.  Then from \eqref{eqn:schr} we have
\bea
\label{eqn:Aeq}
&&-i  \frac{\dot{A_k}(t)}{2}(\chi_k-\Phi_k(t))^2+i A_k(t)(\chi_k-\Phi_k(t))\dot{\Phi_k}(t)- \dot{P_k}(t) \chi_k- \dot{\theta}_k(t)+i\dot{N}_k(t)/N_k(t) \nonumber\\
&& \qquad  =\frac{1}{2a^3}\left[A_k(t)- \left(-A_k(t)(\chi_k-\Phi_k(t))+iP_k(t)\right)^2\right]+\frac{1}{2}a^3K(t) \chi_k^2 \,.
\eea
From  Eq.~\eqref{eqn:hamdot}, $\dot{P_k}(t) =- a^3 K(t)\Phi_k$, so we
can rearrange, 
\bea
&&\frac{1}{2}\left[-i\dot{A_k}(t)+\frac{A_k^2(t)}{a^3}-a^3K(t)\right](\chi_k-\Phi_k)^2\nonumber\\
&&+\frac12 K(t)a^3\Phi_k^2 + i\dot{N}_k(t)/N_k(t)
-\dot{\theta}_k(t)-\frac{1}{2a^3}A_k(t)-\frac{a^3}{2}\dot{\Phi_k}^2=0 \,.
\eea
For this to hold we should require
\be
\label{eqn:aeq}
-i\dot{A_k}(t)+\frac{A_k^2(t)}{a^3}-a^3K(t)=0 \,.
\ee
and then the normalization $N_k(t)$ and the irrelevant phase $\theta_k(t)$
can be determined by integration (see also Ref.~\cite{Boddy:2016zkn}, which derives the same equation using a slightly different method).

Equation~\eqref{eqn:aeq} is a Riccati equation.  We can rewrite $A_k$ in
terms of an unknown function $f_k$,
\be
iA_k=a^3 \frac{\dot f_k}{f_k} \,,
\ee
so that Eq.~(\ref{eqn:aeq}) becomes
\be\label{eqn:EOMf}
\ddot f_k+3H\dot f_k+K(t)f_k = 0
\ee
which is the just the classical equation of motion,
Eq.~\eqref{eqn:EOM1}, with $f_k(t)$ analogous to $\phi_k(t)$
there. Then let us define $g_k(\tau) =a f_k(\tau)$ analogous to
$\sigma_k(\tau)$ in Eq.~\eqref{eqn:EOMsigma}, with the general solution
given by  Eq.~(\ref{eqn:classsolJY}),
\be\label{eqn:gsol}
g_k = \sqrt{-\tau}\left[g_J J_\nu(-k\tau) + g_Y Y_\nu(-k\tau)\right] \,.
\ee
Then we have
\be
A_k = -i a^3\left[\frac{g_k'}{g_ka}-H\right]
=i a^2 k\left[\frac{g_J J_{\nu-1}(-k\tau)+ g_Y Y_{\nu-1}(-k\tau)}
{g_J J_\nu (-k\tau)+ g_Y Y_\nu(-k\tau)}
+\frac{\epsilon Ha}{2k}\right] \,.
\ee

The precision (inverse variance) of the Gaussian wave function is
given by
\be
\label{eqn:Regen}
\Re A_k = a^2\Im\frac{g_k'}{g_k} =a^2\frac{g_k^*g_k' - g_k'^*g_k}{2i|g_k|^2} \,.
\ee
The numerator of \eqref{eqn:Regen} is the Wronskian of the solutions
$g_k$ and $g_k^*$.  It does not depend on time and its real part
vanishes.  To have a sensible wavefunction, we need $\Re A_k > 0$, so
$\Im (g_k^*g_k' - g_k'^*g_k)>0$.  We have the freedom to multiply
$g_k$ by a constant without changing $A_k$, and we can use this
freedom to require
\be\label{eqn:Wronskian}
g_k^*g_k' - g_k'^*g_k= i\,.
\ee
or if $g_k$ has the form of Eq.~(\ref{eqn:gsol}), 
\be\label{eqn:gconstraint}
\Im(g_J g_Y^*) = \pi/4 \,.
\ee
When the parameters of Eq.~(\ref{eqn:gsol}) obey
Eq.~(\ref{eqn:gconstraint}), we have the simple form
\be\label{eqn:realA}
\Re A_k = \frac{a^2}{2|g_k|^2}
= -\frac{a^2}{2\tau \left|g_J J_\nu(-k\tau)+ g_Y Y_\nu(-k\tau)\right|^2} \,.
\ee

For $\Im A_k$, we take the imaginary part of  Eq.~\eqref{eqn:aeq},
\be
-\Re \dot A_k + \frac{2 (\Re A_k)(\Im A_k)}{a^3}= 0
\ee
so
\be
\Im A_k = \frac{a^3}{2}\frac{d}{dt}\ln(\Re A_k) \,.
\ee
The imaginary part of $A_k$ tells us how the width of the Gaussian is changing.

At late times, $J_\nu$ decreases while $Y_\nu$ increases.  Thus most
properties depend on $g_Y$.  We can use the freedom to multiply
$g_J$ and $g_Y$ by a common phase without affecting $A$ or
Eq.~(\ref{eqn:gconstraint}) to make $g_Y$ purely imaginary and write
the general $g_k$ satisfying Eq.~(\ref{eqn:gconstraint}) in the form
\be\label{eqn:gall}
g_k(\tau)=(1/2)\sqrt{-\pi \tau/c}((1+ib)J_\nu (-k\tau)-i c Y_\nu(-k\tau)) \,,
\ee
where $c$ is a real constant giving the degree of squeezing.

From Eq.~(\ref{eqn:gall}), we find
\be\label{eqn:Ageneral}
A_k =i a^2 k\left[\frac{(1+ib) J_{\nu-1}(-k\tau)-ic Y_{\nu-1}(-k\tau)}
{(1+ib) J_\nu (-k\tau)-ic Y_\nu(-k\tau)}
+\frac{\epsilon Ha}{2k}\right]
\ee
and
\be\label{eqn:realAgeneral}
\Re A_k = \frac{2a^3Hc}{\pi
\left(J_\nu(-k\tau)^2 + \left(c Y_\nu(-k\tau)-b J_\nu(-k\tau)\right)^2\right)}\,.
\ee

\subsection{Late and early time approximations}

At late times, we can neglect $J_\nu$ and then use the small-argument
approximation for $Y_\nu$,
\be\label{eqn:realAlate}
\Re A_k = \frac{2a^3H}{\pi c Y_\nu(-k\tau)^2}
= \frac{\pi k^3}{4c\Gamma(\nu)^2 H^2}
\left(-\frac{2}{k\tau}\right)^{2\epsilon}\,.
\ee
Note that $b$ does not enter, because it is the coefficient of a
decaying mode.   The width (standard deviation) of the Gaussian wavefunction is
\be\label{eqn:s}
s=\frac{1}{\sqrt{\Re A_k(\tau)}} =2\sqrt{\frac{c}{\pi}} \frac{\Gamma(\nu)H}{k^{3/2}} \left(-\frac{2}{k\tau}  \right)^{-\epsilon} \,.
\ee
At late times, the imaginary part of $A$ becomes
\be
\label{eqn:imAlate}
\Im A_k(\tau) = a^2 k\left[\frac{Y_{\nu-1}(-k\tau)}{Y_\nu(-k\tau)}
+\frac{\epsilon Ha}{2k}\right]
= \frac{ k^2a}{2 H (\nu-1)}+\epsilon a^3 H \,.
\ee
Thus we see that $\Im A_k$ grows rapidly with time,
so the phase of $\psi_k(\phi_k)$ oscillates rapidly.

At early times we can take
\bea
\sqrt{-\tau} J_\nu(-k\tau) &=& \sqrt{\frac{2}{\pi k}}\cos(k\tau +\delta_\nu)\\
\sqrt{-\tau} Y_\nu(-k\tau) &=& -\sqrt{\frac{2}{\pi k}}\sin(k\tau +\delta_\nu)
\eea
Combining these with any complex coefficients $g_J$ and $g_Y$ gives an
ellipse in the complex plane.   Equation~(\ref{eqn:gconstraint}) tells us
that the area of this ellipse is $\pi/(2 k)$.  We can always write it
\be\label{eqn:gearly}
g=\frac{e^{i\theta}}{\sqrt{2\bar ck}}\left[\cos(-k\tau+\delta)- i\bar c \sin(-k\tau+\delta)\right]
\ee
with $\bar c$, $\theta$, $\delta$ real constants that depend on $c$
and $b$.  Without loss of generality we can take $\bar c \le 1$.
Equation \eqref{eqn:gearly} gives
\be
\Re A = \frac{a^2k\bar c}{\cos^2(-k\tau+\delta)+ \bar c^2\sin^2(-k\tau+\delta)}\,,
\ee
which is an oscillation between $\bar c$ and $1/\bar c$, with an overall
factor $a^2 k$, which is the $A$ of the ground state.

If $b=0$, Eq.~(\ref{eqn:gearly}) is
just the early-time version of Eq.~(\ref{eqn:gall}) with $c = \bar
c$.  Otherwise, $\bar c< c$, i.e., some of the early oscillation in
$\Re A$ is damped rather than frozen in.

All of the considerations above apply also to the wave function
$\zeta_k(\xi_k)$, which could in principle have its own $A_k$,
$\Phi_k$ and $P_k$.  An even more general Gaussian would have a
general quadratic form in $\chi_k$ and $\xi_k$ in the exponent, i.e.,
there would be an additional term proportional to $\chi_k \xi_k$.
However, in the cases we will use, the $A_k$ will be the same for both
$\chi_k$ and $\xi_k$, and there will be no cross term.  This means
that considerations involving the width of the Gaussian will not
distinguish particular locations $\bx$.

The values of $\Phi_k$ and $P_k$ may be different for $\chi_k$ and
$\xi_k$, but this means merely that we can build our wavefunction
around any classical solution such as Eq.~(\ref{eqn:classsolJY-}).

\subsection{Bunch-Davies vacuum}

Now we will consider the effect of fluctuations of $\phi$ in the
background de Sitter spacetime.  After long period of inflation, we
expect the quantum state to be the Bunch-Davies vacuum
\cite{Bunch:1978yq}.  Consider a short-wavelength mode, so $k\gg H\gg
m$, and the mode is at each moment in the flat-space ground state,
which is a Gaussian around $\phi = 0$ (so $\Phi_k = P_k = 0$) with
$A_k =a^2 k$.  From Eq.~(\ref{eqn:realA}), this requires
$|g_k|=1/\sqrt{2k}$.  Thus at early times we have
Eq.~(\ref{eqn:gearly}) with $\bar c = 1$ (and we can take $\theta =
0$), $g_k = (1/\sqrt{2k})e^{ik\tau}$.  This becomes in general
Eq.~(\ref{eqn:gall}) with $c = 1$ and $b = 0$, i.e., $g_k(\tau)=
(\sqrt{-\pi\tau}/2) H^{(2)}_\nu(-k\tau)$ meaning $g_J = \sqrt{\pi}/2$,
$g_Y = -i\sqrt{\pi}/2$ and so
\be\label{eqn:BDrealA}
\Re A_k=  \frac{2a^3H}{\pi\left(J_\nu(-k\tau)^2+ Y_\nu(-k\tau)^2\right)}\,.
\ee
At late times,
\be\label{eqn:BDrealAlate}
\Re A_k = \frac{\pi k^3}{4\Gamma(\nu)^2 H^2}
\left(-\frac{2}{k\tau}\right)^{2\epsilon}\,,
\ee
and the width of the wavefunction is
\be\label{eqn:s0}
s_0=\frac{2}{\sqrt{\pi}} \frac{\Gamma(\nu)H}{k^{3/2}} \left(-\frac{k\tau}{2}  \right)^\epsilon \,.
\ee

In the massless case, $\epsilon = 0$, the width is fixed.  The width
the of probability distribution $|\psi|^2\sim e^{-\Re A\phi^2}$ is
$s/\sqrt{2}$.  Using $\Gamma(3/2)=\sqrt{\pi}/2$, we recover the usual
perturbation spectrum from inflation, $\delta_k=H/\sqrt{2k^3}$.  With
a mass the Gaussian narrows with time.

\subsection{State with rapid expansion}

When the Bunch-Davies state above is strongly subhorizon, it is
essentially in its ground state.  The rate the frequency changes is
small compared to the frequency itself, so conditions change
adiabatically and the mode remains in the ground state.  But as it
crosses the horizon, this is no longer true.  The mode remains a
Gaussian whose width is given by Eq.~(\ref{eqn:s0}), but it is no
longer an energy eigenstate.  At late times when the mode is
superhorizon, it is frozen in and its energy depends almost entirely
on $\phi$.  We can think of this mode as having a spread of possible
energies corresponding to different $\phi$.

We now imagine that we have extended our toy model to include some
small interactions between different modes, such as nonlinear
gravitational coupling \cite{Nelson:2016kjm}, which will lead to
decoherence.  Each (sufficiently superhorizon) mode wave function will
decohere into some particular alternative narrowly peaked around some
$\Phi_k$, with the chance of the various $\Phi_k$ given by
\be\label{eqn:Phiprobability}
|\psi_k(\Phi_k)|^2 \sim e^{-(\Re A_k )\Phi_k^2}
\ee
with $\Re A_k$ given by Eq.~(\ref{eqn:BDrealAlate}).  More specifically,
decoherence at late times will pick out a specific classical solution
of the slowly rolling type given by Eq.~(\ref{eqn:slowmode}) above. 

The details of the decoherence will not concern us here, but it is
important that the different modes decohere independently and that
the ``pointer basis'' of decoherence is the space of field values.
This latter fact is required by the unobservability of the decaying
mode and the rapid phase oscillations of $\psi$
\cite{Polarski:1995jg,Kiefer:1998qe,Burgess:2014eoa,Nelson:2016kjm}.

It is important to note that the probability distribution of
Eq.~(\ref{eqn:Phiprobability}) extends to values of $\Phi_k$ that will
have much more energy than the average in the Bunch-Davies state.
Indeed, such values are the ones that will lead to eternal inflation.
This increase in the (average) energy is possible even though the
decoherence process (weak coupling to other modes
\cite{Nelson:2016kjm}, for example) does not transfer any significant
energy into the mode.  Since the Bunch-Davies state of a
superhorizon mode is not an eigenstate of the Hamiltonian, decoherence
without energy transfer can select a state with energy different from
the average energy of the state before decoherence.    The decoherence
process is analogous to the final measurement of the system of
Sec.~\ref{sub:smallchanges}.

After decoherence, the mode of interest will have only some small
spread of $\phi$.  Let us take it to be a Gaussian with narrow width
$s<s_0$.  We will take $s$ to be the same for the real and imaginary
components ($\chi_k$ and $\xi_k$) of $\phi_k$, as we discuss below.
We would now like to know the history of that specific state, to see
explicitly whether it obeys energy conditions.

The general Gaussian state with width $s$ is given by
Eqs.~(\ref{eqn:gall},\ref{eqn:s}).  The ratio $s/s_0 = \sqrt{c}$ is a
constant less than 1, which gives the degree of decoherence, i.e., it
tells us how narrow the decoherent wave function is as compared to the
wave function before decoherence.  We cannot determine $b$, because it
does not enter at late times.  The expressions for $g$ and $A$ are
given in Eq.~(\ref{eqn:gall}--\ref{eqn:imAlate}).

\section{Energy conditions}
\label{sec:energy}

The stress-energy tensor for the minimally coupled scalar field is
\be\label{eqn:Tmn}
T_{\mu \nu}=(\partial_\mu \phi \partial_\nu \phi)-\frac{1}{2} g_{\mu \nu} ((\nabla \phi)^2-m^2\phi^2) \,.
\ee
We will be interested in the expectation value of $T_{\mu \nu}$ in
quantum states given by products of wavefunctions in the form of
Eq.~(\ref{eqn:state}).  Such a state has a classical part
\be
\Phi = \frac{1}{L^{3/2}}\sum_{\bk} e^{i\bk\cdot\bx} \Phi_k\,,
\ee
and we can define the purely quantum part, $\delta\phi =\phi-\Phi$,
with the usual Fourier decomposition
\be
\delta\phi = \frac{1}{L^{3/2}}\sum_{\bk} e^{i\bk\cdot\bx} \delta\phi_k
\ee
and write $\delta\phi_k = \delta\chi_k + i \delta\xi_k$.  Then our
wavefunction $\Psi$ is the product of terms
\be
\psi_k(\delta\chi_k) \sim
\exp{\left[-\frac{A_k(t)}{2}\delta\chi_k^2+iP(t)(\delta\chi_k+\Phi_k)\right]} \,.
\ee
and similarly for $\delta\xi_k$.

Using $\phi=\delta\phi +\Phi$ in Eq.~(\ref{eqn:Tmn}), we see that
$T_{\mu\nu}$ is made up of quadratic combinations of the operators
$\chi_k$, $\xi_k$, and their corresponding momenta.  Applying such
operators to the wavefunction $\Psi$, we get expectation values of
terms up to second order in the $\delta\chi_k$ and $\delta\xi_k$.  Terms
with exactly one of these vanish since
$\langle\delta\chi_k\rangle=\langle\delta\xi_k\rangle = 0$ by symmetry.
Terms with no $\delta\chi_k$ or $\delta\xi_k$ are are just classical
quantities, while those with two are quantum mechanical, but they
vanish unless the $k$'s are the same and they are both $\delta\chi_k$ or
both $\delta\xi_k$.  Thus we can write $T_{\mu\nu}$ as the sum of a
classical and a quantum part,
\be
T_{\mu\nu} = T_{\mu\nu}^C + T_{\mu\nu}^Q
\ee
with
\be\label{eqn:TmnC}
T_{\mu\nu}^C=(\partial_\mu \Phi \partial_\nu \Phi)-\frac{1}{2} g_{\mu
  \nu} ((\nabla \Phi)^2-m^2\Phi^2)\,,
\ee
and
\be\label{eqn:TmnQ}
T_{\mu\nu}^Q=(\partial_\mu \delta\phi \partial_\nu \delta\phi)-\frac{1}{2} g_{\mu \nu} ((\nabla \delta\phi)^2-m^2\delta\phi^2) \,.
\ee
The classical part obeys the energy conditions, while the quantum
mechanical part may violate them.

Modes which are subhorizon (or those which are super horizon but not
sufficiently so to have decohered \cite{Nelson:2016kjm}) are still in
the Bunch-Davies state.  They have no classical part and their $\Re A$ is
given by Eq.~(\ref{eqn:BDrealA}).  Decoherent modes have classical
parts around which their wave functions are strongly peaked, so their
$\Re A$ is much larger than that of Bunch-Davies.

Since the overall stress-energy tensor is just the sum of quantum and
classical parts, we will first calculate the energy density and the
NEC in the case without any classical contribution, so $T_{\mu\nu}$ is
just $T_{\mu\nu}^Q$ and $\delta\phi$ is just $\phi$.  Then we can add
in a possible classical contribution later.  We will assume that the
decoherent $A_k$ does not depend on the particular $\Phi_k$ that
decoherence gives us, i.e., the degree of decoherence is the same in
the different possible decoherent states.  This is consistent with
Ref.~\cite{Nelson:2016kjm} where the decoherence functional depends
only on the difference between its two arguments.  In other words,
$\psi_k(\phi_k)$ is independent of $\Phi_k$ and $P_k$.  Thus $\psi_k(\phi)$
depends only on the magnitude of $k$ and not on its direction, and the
real and imaginary parts of $\phi$ enter in the same way.  Then
$T_{\mu\nu}$ has the perfect fluid form, $\diag(\rho, P, P, P)$.
The projection on any null vector with unit time component $n$ is then
\be
T_{\mu\nu}n^\mu n^\nu =\rho +P\equiv N \,.
\ee

The energy density is
\be
\rho=T_{00}=\frac{1}{2}(\partial_0 \phi)^2+\frac{1}{2} \left(\frac{1}{a^2}(\partial_i \phi)^2+m^2\phi^2\right) \,,
\ee
and the null projection
\be
N = T_{00}+\frac{1}{3}{T}_i^i=(\partial_0\phi)^2+\frac{1}{3a^2} (\partial_i \phi )^2 \,.
\ee
When we decompose the field into modes, we get
\be
\rho=\frac{1}{2V}\sum_k \left[|\dot\phi_k|^2+\left(\frac{k^2}{a^2}+m^2\right)|\phi_k|^2\right]=\frac{1}{2a^2V} \sum_k \left[|\phi_k'|^2 +\left(k^2+a^2 m^2 \right)|\phi_k|^2  \right]  \,,
\ee
and
\be
N=\frac{1}{V} \sum_k \left[|\dot\phi|^2_k+\frac{k^2}{3 a^2}|\phi_k|^2\right]= \frac{1}{a^2V} \sum_k \left[|\phi_k'|^2+\frac{k^2}{3} |\phi_k|^2 \right]   \,.
\ee
When we quantize the field, $\rho$ and $N$ become operators, which
we can write in terms of the operators $\chi_k$ and $\xi_k$, being the
real and imaginary parts of $\phi_k$, and their conjugate momenta,
\bea
\rho&=&\frac{1}{2a^2V} \sum_k \left[\frac{p_k^2}{a^4} +\left(k^2+a^2
  m^2 \right)\chi_k^2  \right] + \text{terms involving $\xi_k$} \,, \\
N&=&\frac{1}{a^2V}\sum_k \left(\frac{p_k^2}{a^4}+\frac{k^2}{3} \chi_k^2  \right)
+ \text{terms involving $\xi_k$} \,.
\eea
Next we want to find the expectation value of the quantized energy
density in a squeezed coherent state of the form of
Eq.~\eqref{eqn:state}.
We first calculate 
\be
\langle \Psi | \chi_k^2 | \Psi \rangle=\int_{-\infty}^\infty d\chi_k\,
\,\chi_k^2 |\Psi(\phi,\tau)|^2=\frac{1}{2 \Re(A_k(\tau))}\,,
\ee
\be
\langle \Psi |  p_k^2 | \Psi \rangle=\int_{-\infty}^\infty d\chi_k\,  \left|\frac{\partial}{\partial \phi} \Psi(\phi,\tau)\right|^2=\frac{|A_k(\tau)|^2}{2 \Re(A_k(\tau))} \,.
\ee
The contribution for a single mode of $\chi_k$ is then
\be
\label{eqn:rhomode}
\langle \Psi |\rho_k | \Psi \rangle = \frac{1}{2a^2V} \left[\frac1{a^4}\left(\frac{|A_k(\tau)|^2}{2\Re(A_k(\tau))}
\right) +\left(k^2+a^2 m^2\right)\left(\frac1{2\Re(A_k(\tau))}\right)\right] \,,
\ee
and
\be
\langle \Psi | N_k | \Psi \rangle=\frac{1}{a^2V}
\bigg[\frac1{a^4}\left(\frac{|A_k(\tau)|^2}{2\Re(A_k(\tau))}
\right) +\frac{k^2}{3} \left(\frac1{2\Re(A_k(\tau))}\right)\bigg] \,.
\ee

To better understand these formulas, let us define a dimensionless
quantity 
\be
B = \frac{1}{ka^2} A_k(\tau) \,.
\ee
From Eq.~(\ref{eqn:Ageneral}), we find that $B$ depends on $k$ and
$\tau$ only through the combination $k\tau$ and through the possible dependence
(absent in the Bunch-Davies state) of the parameters $c$ and $b$ on
$k$.

If we consider the massless case $m=0$ for simplicity, we have
\be\label{eqn:rhoB}
\langle \Psi |\rho_k | \Psi \rangle =\frac{H^4}{V k^3}\frac{(k\tau)^4}{4} \left(\frac{|B|^2}{\Re{B}}+\frac{1}{\Re{B}} \right) \,.
\ee
and
\be\label{eqn:NB}
\langle \Psi | N | \Psi \rangle=\frac{H^4}{V k^3}\frac{(k\tau)^4}{2}  \left( \frac{|B|^2}{\Re{B}}+\frac{1}{3}\frac{1}{\Re{B}}\right) \,.
\ee
We note that $H^4 (k\tau)^4/(Vk^3)=k_{\text{phys}}/V_{\text{phys}}$
where $k_{\text{phys}}=k/a$ and $V_{\text{phys}}=Va^3$ are the
physical wavenumber and volume respectively. So 
Eqs.~(\ref{eqn:rhoB},\ref{eqn:NB}) are physical quantities as they
should be.

\subsection{Renormalization}

While expectation values in single modes, such as
Eqs.~(\ref{eqn:rhoB},\ref{eqn:NB}), are finite, sums over $k$ lead to
the usual divergence of the expectation value of the
stress-energy tensor in quantum field theory.  To get finite physical
values we must renormalize.  There is a well-known prescription
following a set of axioms described by Wald \cite{Wald:1978pj}.  We
first write a ``point-split'' version of $T_{\mu\nu}$ as a
differential operator acting on the two-point function.  We
renormalize the two-point function by subtracting the Hadamard
parametrix, yielding a smooth function.  We then apply the
differential operator and bring the points together.  One can use this
procedure \cite{Hack:2015zwa,Bunch:1978yq,Eltzner:2010nx} to compute
$T_{\mu\nu}$ in the Bunch-Davies vacuum and in a general Gaussian
state.

However, in our case there is an additional complexity. We are
interested in the energy density and null-projection of a single mode.
Thus we would need to renormalize not just the overall $\rho$ and $N$,
but the contribution from each mode separately.  This requires a mode
expansion of the Hadamard parametrix, but that is not a clearly
defined operation.  (For an analysis of the corresponding problem for
the Casimir effect in flat space, see Ref.~\cite{Ford:1988gt}.)

For this reason, we will not try to give the contribution of a single
mode to the renormalized $T_{\mu\nu}$.  Instead we will give the
difference $T_{\mu\nu}^{\text{diff}}$ between the contribution of a mode in a
particular state and that same mode in the Bunch-Davies vacuum.  This
is unambiguous. \footnote{Alternatively one could perhaps use here the adiabatic regularization process \cite{birrell1978application} to renormalize mode-by-mode. This is a point that deserves further investigation}

The Bunch-Davies vacuum obeys the symmetries of de Sitter space, which
means that its renormalized $T_{\mu\nu}$ (including all modes) must be
proportional to $g_{\mu\nu}$ \cite{Bunch:1978yq}.  Thus its NEC
contribution is zero, and it marginally obeys NEC.

Thus we write the total renormalized energy density in any state as
the positive Bunch-Davies (BD) value plus the sum of $\rhodiff_k =
\rho_k - \rho_k^{BD}$.  The total contribution to $N$ is just the sum
of of $\Ndiff_k = N_k - N_k^{BD}$, since the total $N = 0$ in the
Bunch-Davies state.

Following that idea we subtract the Bunch-Davies vacuum which corresponds to the $c=1$ case 
\be
\langle \Psi |\rhodiff | \Psi \rangle =\langle \Psi |\rho | \Psi \rangle-\langle \Psi |\rho_{c=1} | \Psi \rangle \,.
\ee
For early times we can use Eq.~\eqref{eqn:gearly} which for $c=1$ gives $B=1$. Then
\be
\langle \Psi |\rhodiff | \Psi \rangle =\frac{H^4}{V k^3}\frac{(k\tau)^4}{4}  \left(\frac{|B|^2}{2\Re{B}}+\frac{1}{2\Re{B}}-1 \right)=\frac{H^4}{V k^3}\frac{(k\tau)^4}{4}  \frac{1}{\Re{B}} \left(\Im{B}^2+(\Re{B}-1)^2 \right) \,.
\ee
Since $\Re{B} \geq 0$ the change to the energy density is always
positive: A mode in a narrower Gaussian leads to higher energy density
than the Bunch-Davies state.  However, at late times we can have
decreases.

For the case of $N$, we start with Eq.~(\ref{eqn:NB}) and subtract the
Bunch-Davies vacuum.  At early
times we have
\be
\langle \Psi |\Ndiff | \Psi \rangle=\frac{H^4}{V k^3}\frac{(k\tau)^4}{6}  \left( \frac{3|B|^2}{\Re{B}}+\frac{1}{\Re{B}}-4\right)
\ee
but this does not have a uniform sign, so the NEC contribution of this
mode may be less than in Bunch-Davies.  Since NEC is obeyed only
marginally in Bunch-Davies, this means that some decoherent states
will have NEC violation.

However, in addition to the quantum mechanical contribution, there
will be a classical contribution.  The typical form for the classical
$\Phi$ is one whose amplitude (not including any decaying mode)
matches the standard deviation of the probability distribution $|\psi_k|^2$
in Bunch-Davies state.  This
has the general form (we have chosen the case where $\Phi_k > 0$ at
late times),
\be  
\Phi_k(\tau)=-\frac{\sqrt{\pi}}{2 a}\sqrt{-\tau}Y_\nu(-k\tau)
\ee
or in the massless case,
\be  
\Phi_k(\tau)=\frac{H}{\sqrt{2 k^3}}\left[\cos{(k\tau)}+k \tau \sin{(k\tau)}\right] \,.
\ee
The classical contribution to $\rho$ and $N$ is always positive and for typical values outweighs any negative quantum contribution.  To have eternal inflation, we need an especially large classical part, so this will be even more true in that case.

\subsection{Numerical evolution}

Now we explore these effects numerically. We vary $c$, $b$ and $\nu$
and examine the effect that each parameter has on the change to the
expectation values of $\rhodiff$ and $\Ndiff$.  We will plot
these quantities in units of $H^4/(Vk^3)$.

\begin{figure}
	\includegraphics[scale=0.5]{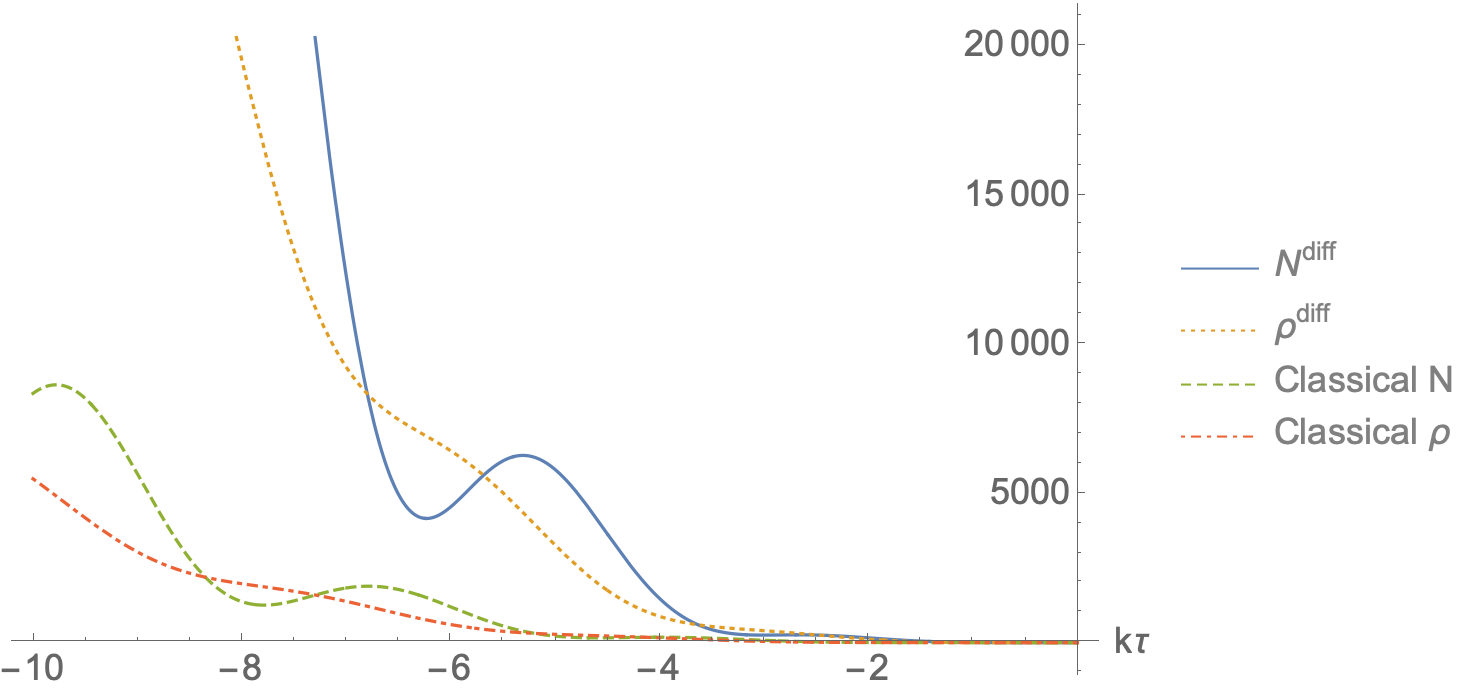}
	\caption{The changes to energy density and (directionally averaged) NEC contribution,
		$\rhodiff$ and $\Ndiff$, from a single mode with $c=0.05, b=0$, and
          the classical part in the massless case. The vertical scale is in units of
          $H^4/(Vk^3)$.  The energy density and NEC diverge at early
          times.}
	\label{fig:005all}
\end{figure}

\begin{figure}
	\includegraphics[scale=0.5]{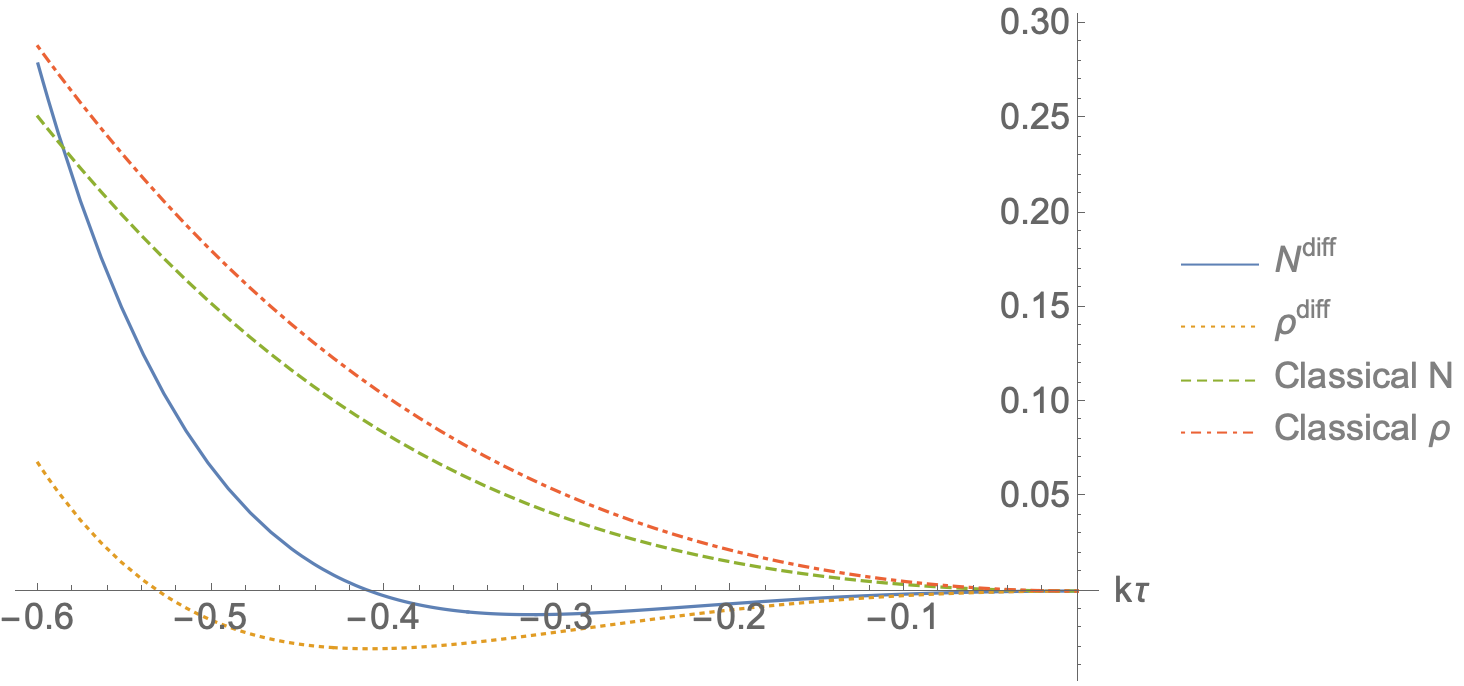}
	\caption{A zoomed in part of Fig.~\ref{fig:005all} where the
          negative values of $\rhodiff$ and $\Ndiff$ are
          visible.}
	\label{fig:005late}
\end{figure}

First we look at $m=0$ and small values of $c$, which correspond to
highly decoherent states. In Fig.~\ref{fig:005all}
we see that at early times $\rhodiff$, $\Ndiff$, and of
course the classical part, are positive.  Thus the selection of a
decoherent state does not lead to negative energies or NEC violation.
All effects are larger at earlier times. However, at late times
(Fig.~\ref{fig:005late})
there are small negative values of both $\rhodiff$ and
$\Ndiff$. Still, a typical classical part is expected to be
larger. A numerical analysis shows no decrease in the values of the
NEC or the energy density when both the classical and quantum parts
are taken into account.  So a typical decoherent state does not
produce negative energies or NEC violation even at late times.

\begin{figure}
  \includegraphics[scale=0.33]{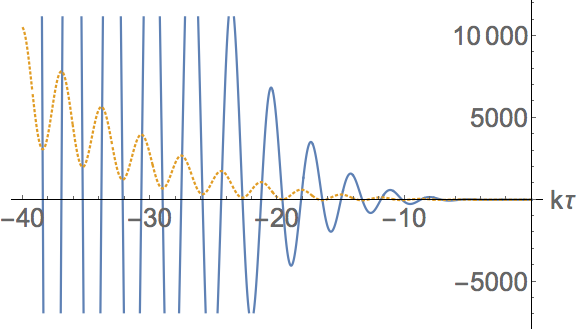}
  \includegraphics[scale=0.33]{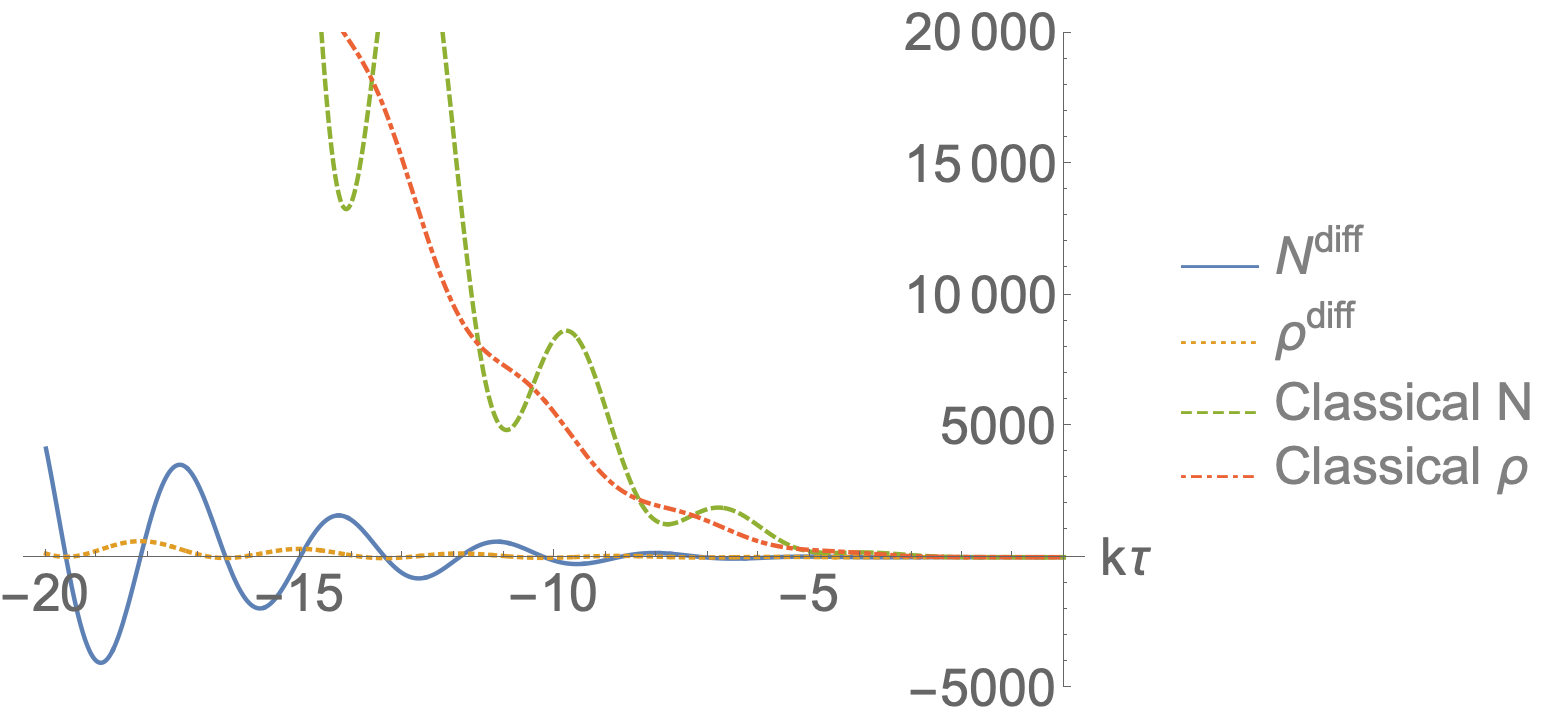}
   \caption{As in Figs.~\ref{fig:005all} and \ref{fig:005late}, but for
     $c=0.9$. Here $\rhodiff$ is positive at early times while
     $\Ndiff$ oscillates (left). But both are much smaller than their
     classical counterparts at early times (right).}
	\label{fig:09}
\end{figure}

\begin{figure}
	\includegraphics[scale=0.5]{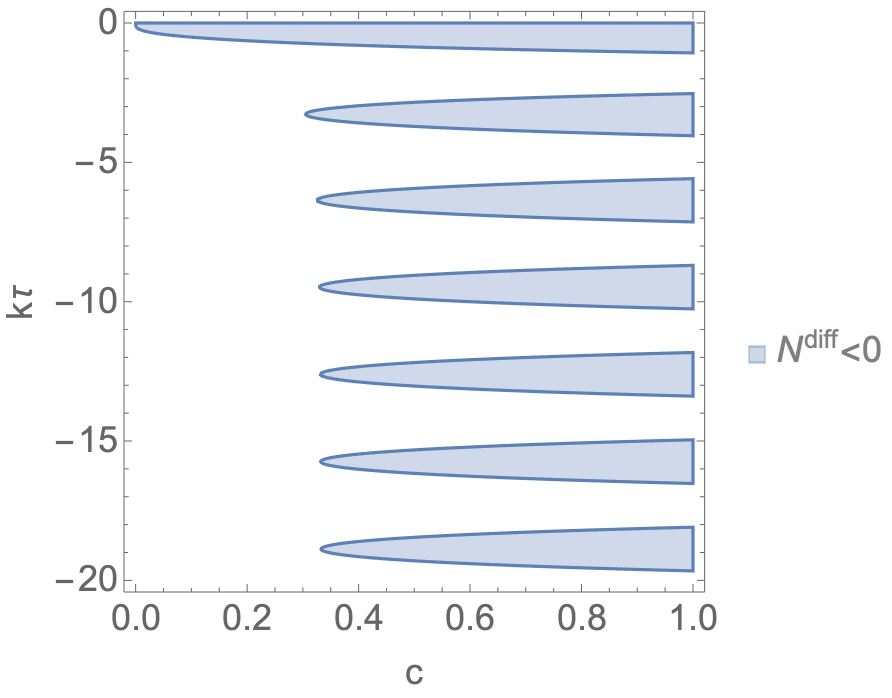}
	\caption{Plot of the regions where $\Ndiff<0$.   For
          small $c$ this occurs only at late times, while for larger
          $c$ it occurs repeatedly, with longer duration for $c$
          closer to 1.}
	\label{fig:region}
\end{figure}

We can ask what happens for different values of $c$. For large values
of $c$ (close to 1) both $\Ndiff$ and $\rhodiff$ have recurring
negative values.  At early times, $\rhodiff$ becomes strictly
positive as expected, but that doesn't happen with
$\Ndiff$. However, the classical values are much larger at early times
in both cases (Fig.~\ref{fig:09}).
In the regional plot of Fig.~\ref{fig:region} 
we see that these recurring regions of negative $\Ndiff$
become longer for $c$ closer to $1$.

\begin{figure}
	\includegraphics[scale=0.35]{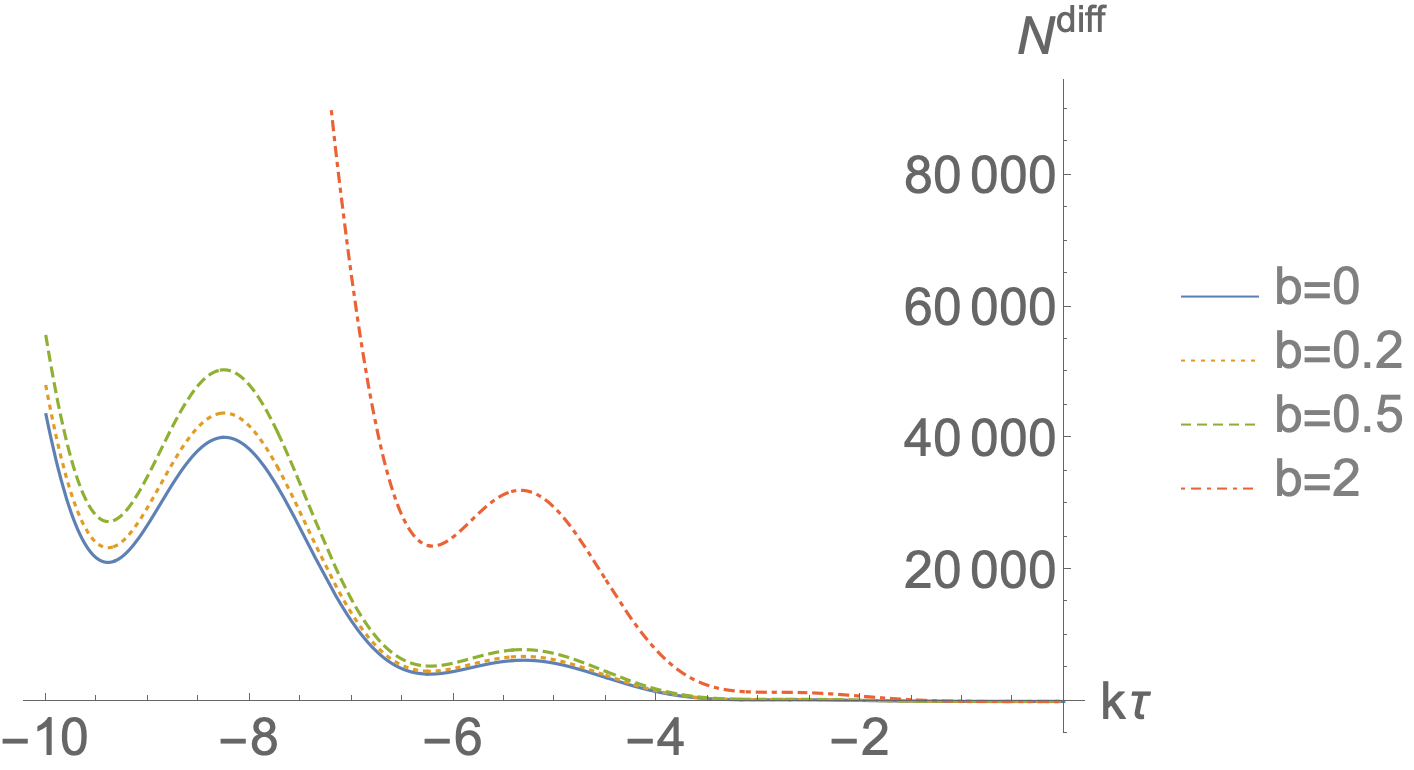}
	\includegraphics[scale=0.35]{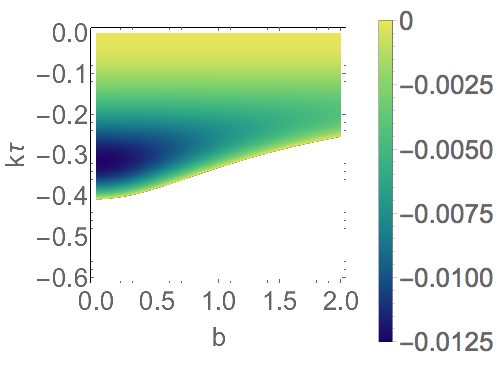}
	\caption{The effect of parameter $b$ on
          $\Ndiff$. It becomes more
          positive with increasing $b$ (left) while at late times
          it is less negative and negative for a shorter time (right).}
	\label{fig:b}
\end{figure}

The effect of the $b$ parameter, the coefficient of the decaying mode,
is shown in Fig.~\ref{fig:b}.
For small values of $c$, increasing the $b$ parameter reduces the
range of over which $\Ndiff<0$ and the maximum negative magnitude that it reaches.

\begin{figure}
	\includegraphics[scale=0.5]{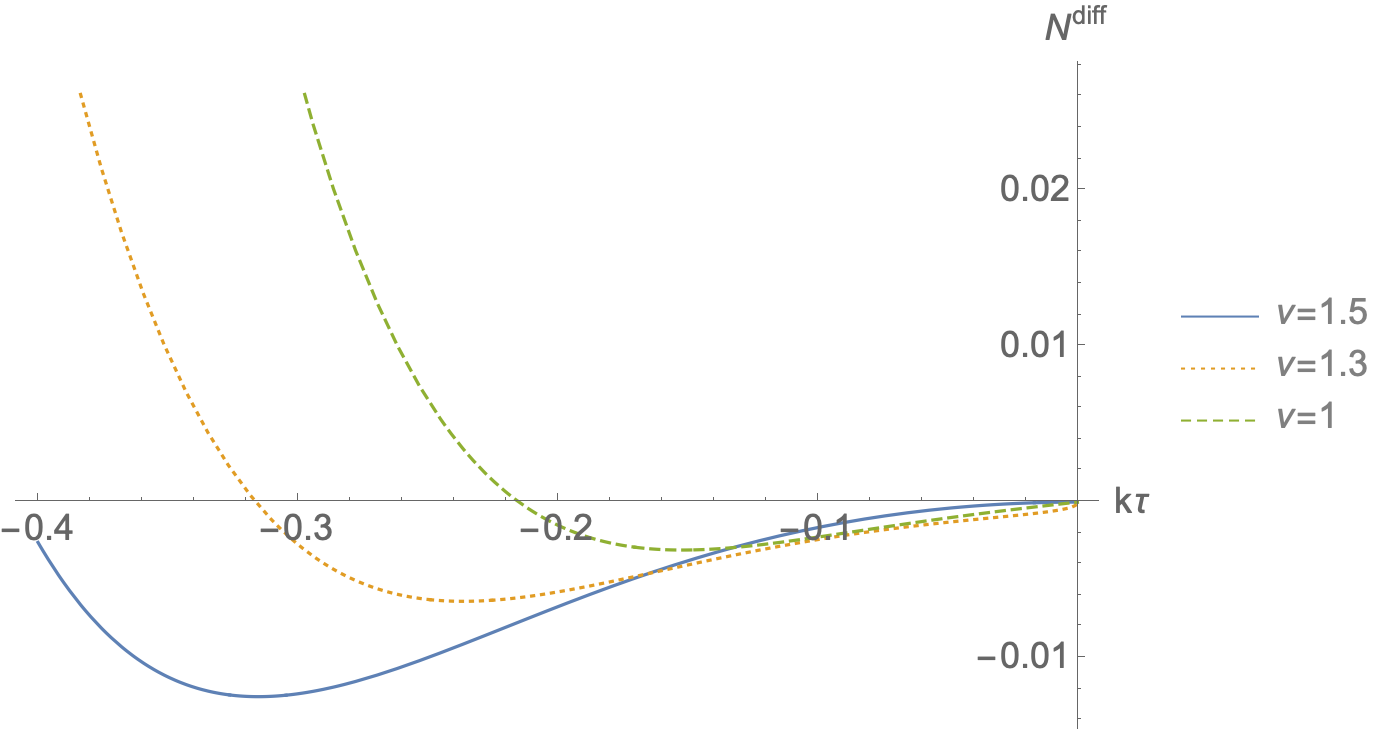}
	\caption{The effect of the mass on $\Ndiff$. For decreasing values of the $\nu$ parameter (increasing mass) the negative values of $\Ndiff$ at late times decrease in duration and magnitude.}
	\label{fig:mass}
\end{figure}

Finally we can examine the effect of mass on the NEC. For a larger
mass, according to Eq.~\eqref{eqn:nu} the $\nu$ parameter of the Bessel
function is something smaller than $3/2$. As we see in
Fig.~\eqref{fig:mass}
larger mass leads to less negative values of the $\Ndiff$ at late times for
small values of $c$.

In every case, a decoherent state with a typical or larger classical
part has a larger energy density and a more positive contribution to
NEC at all times than the Bunch-Davies state.  Thus the sequence of
decoherent states that leads to eternal inflation always obeys NEC.
The process that allows eternal inflation to take place is successive
selection of subsectors of the wavefunction, each of which
individually always obeys NEC.

\section{Beyond the simple model}
\label{sec:realistic}

In the sections above we discussed a simple model that exhibits the
paradox of eternal inflation and energy conditions and showed how this
paradox is resolved.  It seems clear that the same resolution applies
in realistic models, but it would still be good to go beyond the
simple model and discuss the situation of the actual eternal
inflation.  However, this is very complicated.  It is straightforward
to consider a quasi-de Sitter background, and we do that in the next
subsection.  Then we discuss how one might go beyond that.

\subsection{Quasi-de Sitter background}

Above we considered a massive scalar field in a de Sitter background.
Here we will consider instead the fluctuations of an inflaton field
rolling in a linear potential, without any quadratic (mass) term.  The
classical slow-roll motion gives rise to a quasi-de Sitter space, and
we consider the fluctuations to evolve only in this background
spacetime, not in the modified spacetime that they themselves produce.
The scale factor for quasi-de Sitter is \cite{Riotto:2002yw}
\be
a\sim \frac{1}{\tau^{1+\varepsilon}}
\ee
where $\varepsilon\ll 1$ is the slow roll parameter of
Eq.~(\ref{eqn:slowroll}).  The motion in this background is given by
Eq.~(\ref{eqn:EOMsigma}), but now
\be
K(\tau) = k^2-\frac{a''}{a}
= k^2 - \frac{1}{\tau^2}\left(2+3\varepsilon\right)
\ee
giving solutions that are qualitatively the same as before with
\be
\nu^2 = \frac94 -3\varepsilon \,.
\ee
instead of Eq.~(\ref{eqn:nu}).

\subsection{Going further}

The next step would be to include modifications to the spacetime as a
result of a fluctuating field.  This is routinely done in the study of
inflationary perturbations.  In that case, one takes the perturbations
to be small and considers only their first order effects.  For example
$T_{\mu\nu}$ is second order in the field, so one would consider only
cross terms with products of the fluctuating field and the
background.  In this case one can also consider gauge (coordinate)
choices that remove the fluctuation in the field and move it into
other sectors.  For example one can choose surfaces on which the field
has its background value.

In our case, however, the fluctuations are not small.  On the
contrary, the effect of the fluctuations that lead to eternal
inflation is larger than the effect of the slowly rolling background.
So the first order expansion does not make sense.  Furthermore, one
cannot absorb field fluctuations into gauge transformations.  No
(monotonic) change of time coordinate, for example, can take an
increasing field and reparameterize it as a decreasing field.
        
The large fluctuations lead to non-linear effects.  The square of the
field enters into the metric, which then appears in the field
equation.  The nonlinear equation means that the field states are no
longer given by harmonic oscillator states such as the Gaussians that
we have used above.  Furthermore, nonlinearities couple the
different Fourier modes, again invalidating the above analysis.

So we conclude that the case of the real eternal inflation is
approximately given by the massless case in the analysis of sections
Secs.~\ref{sec:class}--\ref{sec:energy} above, but we cannot
consistently go beyond that level.  The idea behind the compatibility
of eternal inflation with energy conditions is the same as in our
simple model, but we cannot do a detailed analysis in the realistic
case.

\section{Conclusion}
\label{sec:conclusion}

Eternal inflation driven by a series of quantum fluctuations that
increase the rate of expansion $H$ of the universe in a certain
region.  But the null energy condition (NEC) requires that $\dot
H\le0$, so $H$ can never increase.  Nevertheless, eternal inflation
can proceed without any NEC violation.  The eternally inflating
spacetime is a succession of more and more specific quantum states,
selected by decoherence.  The energy conditions apply to each of these
quantum states individually, so $H$ cannot increase in any of them.
But when decoherence selects a particular sector of the quantum state,
this sector can have a larger $H$ than the overall state.  A
succession of such decoherence events can lead to a repeated increase
in $H$, even though NEC is obeyed and prohibits an increase $H$ in any
state.

We showed this process explicitly in a simplified model that exhibits
the same paradox.  We considered a scalar field in de Sitter space
where the field can affect the spacetime, but the resulting changes do
not affect the field evolution.  We start in the Bunch-Davies vacuum
and allow decoherence to select narrow states in field space for a
particular mode of the field.  When these states have particularly
large values, they lead to more rapid expansion, but the selected
states themselves never violate NEC.

Would it be right to say that in the selected sector the universe has
always been expanding rapidly, and its expansion rate never increases?
Perhaps not.  In the Bunch-Davies vacuum, when the mode of interest is
strongly sub-horizon it is essentially in an energy eigenstate, so
there is little uncertainty about its contribution to the energy
density and thus to the expansion rate.  An energy eigenstate can of
course be expressed as a superposition of non-eigenstates.  If this
state is coupled to the expansion rate, the state with known expansion
rate can be expressed in terms of states with various different
expansion rates.  Nevertheless the superposition itself has a fixed
expansion rate.  Changes in the Hamiltonian as the mode becomes
super-horizon allow decoherence to pick out one of these states from
the superposition.  That is the sense in which the universe has always
been expanding rapidly in the selected state.

\section*{Acknowledgments}

We would like to thank Larry Ford, Ben Freivogel, Hanno Gottschalk,
Alan Guth, Tanmay Vachaspati, and Alex Vilenkin for helpful
conversations.  EAK's contribution to this work is part of a project
that has received funding from the European Union's Horizon 2020
research and innovation programme under the Marie Sk\l odowska-Curie
grant agreement No. 744037 ``QuEST''.

\bibliography{paper}

\end{document}